 \documentclass[]{emulateapj}

\usepackage{amsbsy}

\shorttitle{Lagrangian coherent structures in a nonlinear dynamo}
\shortauthors{Rempel, Chian, \& Brandenburg}

\begin{document}

\title{Lagrangian coherent structures in nonlinear dynamos}


\author{E. L. Rempel\altaffilmark{1,}\altaffilmark{2}}
\email{rempel@ita.br}

\author{A. C.-L. Chian\altaffilmark{1,}\altaffilmark{3,}\altaffilmark{4}}
\email{abraham.chian@gmail.com}

\and

\author{A. Brandenburg\altaffilmark{5,}\altaffilmark{6}}
\email{brandenb@nordita.org}


\altaffiltext{1}{Institute of Aeronautical Technology (ITA), World Institute
for Space Environment Research (WISER), S\~ao Jos\'e dos Campos -- SP 12228--900, Brazil}
\altaffiltext{2}{Department of Applied Mathematics and Theoretical Physics (DAMTP), University of Cambridge, Cambridge CB3 0WA, UK}
\altaffiltext{3}{National Institute for Space Research (INPE), World Institute for 
Space Environment Research (WISER), P.O. Box 515, S\~ao Jos\'e dos Campos -- SP 12227-010, Brazil}
\altaffiltext{4}{California Institute of Technology, Pasadena, CA 91125, USA}
\altaffiltext{5}{NORDITA, AlbaNova University Ctr, Stockholm, Sweden}
\altaffiltext{6}{Department of Astronomy, Stockholm University, SE-10691 Stockholm, Sweden}

\begin{abstract}
Turbulence and chaos play a fundamental role in stellar convective zones through the transport
of particles, energy and momentum, and in fast dynamos, through the stretching, twisting and folding of magnetic flux tubes. 
A particularly revealing way to describe turbulent motions is through the analysis of Lagrangian coherent structures (LCS), which are  
material lines or surfaces that act as transport barriers in the fluid. 
We report the detection of Lagrangian coherent structures in
helical MHD dynamo simulations with scale separation. 
In an ABC--flow, two dynamo regimes,
a propagating coherent mean--field regime and an intermittent regime, are identified as the 
magnetic diffusivity is varied. The sharp 
contrast between the chaotic tangle of attracting and repelling LCS in
both regimes permits a unique analysis of the impact of the magnetic field on 
the velocity field. 
In a second example, LCS reveal
the link between the level of chaotic mixing of the velocity field and the saturation of a large--scale dynamo 
when the magnetic field exceeds the equipartition value.  
\end{abstract}



\maketitle

\section{Introduction}

The equipartition--strength magnetic fields observed in planets and stars 
are the result of a dynamo process, whereby kinetic energy from the motion 
of a conducting fluid is converted into magnetic energy \citep{axelPR}.
Initially, a weak seed magnetic field \textbf{\emph{B}} undergoes a linear growth in the so--called kinematic
dynamo phase 
until \textbf{\emph{B}} is strong enough to impact the fluid velocity \textbf{\emph{u}}.
Eventually, the magnetic energy saturates due to nonlinear effects.
In a fast dynamo, the growth rate is positive and non--vanishing even in the limit where
the magnetic Reynolds number tends to infinity.
It is known that the growth of the magnetic energy in fast dynamos is related to  
the presence of Lagrangian chaos in the velocity field, i.e., scalar quantities
passively advected by the flow (passive scalars) exhibit chaotic motions \citep{childress95,balsara05}. As \textbf{\emph{B}} grows, it may suppress
this chaos due to backreaction in the velocity field via the Lorentz force, leading 
to the nonlinear saturation of the magnetic energy \citep{cattaneo96,zienicke98}.
A comparison between the chaoticity of the velocity field during the growth and saturation phases
of the dynamo was performed by \citet{axel95}.
In this letter we reveal how magnetic fields can affect the transport of passive scalars through the
formation of transport barriers in the velocity field.

When probing turbulent transport of passive scalars, either Eulerian or Lagrangian tools can be employed.
In the Eulerian approach, for a given velocity field, one can solve an advection--diffusion equation for the 
passive scalar concentration from which a turbulent diffusion coefficient can be computed \citep{vincent96}. 
Moreover, instantaneous snapshots of tracer and velocity fields can be used to extract coherent structures such as eddies and filaments 
\citep{isern04}. 
Alternatively, in the Lagrangian approach the dynamics of fluids is studied by following the trajectories
of a large number of fluid elements or tracer particles. The Lagrangian description has been gaining increasing
attention in the past decade, for example in the study of compressible plasmas \citep{schamel04,padberg07}. It has been
suggested that Lagrangian tools are more appropriate to analyze tracer patterns than their Eulerian counterparts,
since they do not rely solely on snapshots of the velocity field, but measure transport properties along
particle trajectories \citep{dovidio09}. 
We adopt the Lagrangian approach to distinguish the transport properties of three--dimensional (3--D) numerical simulations of compressible magnetohydrodynamic (MHD) dynamos. 
We detect Lagrangian coherent structures (LCS), which are material lines and surfaces in the velocity field that act as barriers to particle transport
and have been described as the Lagrangian building blocks of turbulence \citep{mathur07}.
There are two types of LCS formed by distinct groups of fluid particles, one of them attracts other fluid particles 
and the other one repels them.
These barriers have been used to study turbulence and transport in fluids and plasmas through numerical simulations \citep{green07,padberg07}, 
laboratory experiments \citep{voth02,mathur07},  
and observational data in oceans and a wide range of applications \citep{sandulescu07,olascoaga08,peacock2010}.

Two dynamo models with helical forcing and scale separation are used, the ABC--flow \citep{childress95}
and an isotropic flow driven by a force corresponding to plane waves with random phases \citep{axel2001}.
In the ABC--flow, two different dynamo regimes are investigated,
a regime characterized by a robust spatially coherent mean--field and a regime with intermittent switching between coherent and disordered mean--field  states.
Here, the LCS reveal that the topology of transport barriers in the velocity field suffers a dramatic change 
when the magnetic field undergoes the transition from coherent to intermittent dynamo.
In the second dynamo model, the randomly forced flow, we focus on the problem of the nonlinear saturation
of the magnetic energy. We note a large difference between the patterns of transport barriers in
the kinematic and saturated regimes.

An ultimate goal of this project is to interrelate results from the
Eulerian and Lagrangian approaches. For example, within the Eulerian
approach it has been possible to compute mean-field dynamo transport
coefficients and their magnetic quenching behavior using the test-field
method \citep{axel08b}.
The quenching is being interpreted in terms of a competition between
kinetic helicity that results in an $\alpha$ effect and current
helicity that produces a magnetic $\alpha$ effect.
In an inhomogeneous system, the local current helicity distribution
results from a balance accomplished by magnetic helicity fluxes
(see \citet{axelPR}, for a review).
If there is a direct connection between LCS and the suppression of
turbulent transport and, in particular, the $\alpha$ effect, one might
expect a certain correlation in the spatial patterns of these quantities.
As a preparatory first step, we establish here the basic technique
in the case of a homogeneous turbulent dynamo.

\section{Intermittent dynamo in the ABC--flow}

The dynamo model adopted consists of the compressible MHD equations for an isothermal gas, 
as described by \citet{rempel09}. 
The equations are solved with the {\small PENCIL CODE} \footnote{\tt http://pencil-code.googlecode.com/} 
in a box with sides $L = 2\pi$ and periodic boundary conditions, so the
    smallest wavenumber is $k_1=1$. The sound speed is $c_{\rm s}=1$,
    so our time unit is $(c_{\rm s}k_1)^{-1}$ and the unit of viscosity
    $\nu$ and magnetic diffusivity $\eta$ is $c_{\rm s}/k_1$.
We add to the momentum equation
an external forcing given by the ABC (Arnold--Beltrami--Childress) function,
$
\boldsymbol{f}(\boldsymbol{x})=A_{f}/\sqrt{3}
[(\sin k_{f}z+\cos k_{f}y)\hat{\boldsymbol{x}}, (\sin k_{f}x+\cos k_{f}z)\hat{\boldsymbol{y}}, (\sin k_{f}y+\cos k_{f}x)\hat{\boldsymbol{z}}],
$
where $A_f$ is the amplitude and $k_f$ the wavenumber of the forcing function. 
We use $k_f=5$ to obtain a separation between the energy injection scale and 
the scale of the box, and $A_f = 0.1$, which ensures 
a root mean square velocity $u_{\rm rms}=\langle {\boldsymbol{u}}^2 \rangle^{1/2} < 0.4$.
Following \citet{rempel09}, a numerical resolution of $64^3$ mesh points is chosen. 
The kinetic $(Re)$ and magnetic $(Rm)$ Reynolds numbers
are based on the forcing scale, $Re = u_{\rm rms}/\nu k_f$ and $Rm = u_{\rm rms}/\eta k_f$, 
where $\nu$ is the average kinematic viscosity and $\eta$ the constant magnetic diffusivity.

\begin{figure}[b]
 \includegraphics[width=1.0\columnwidth]{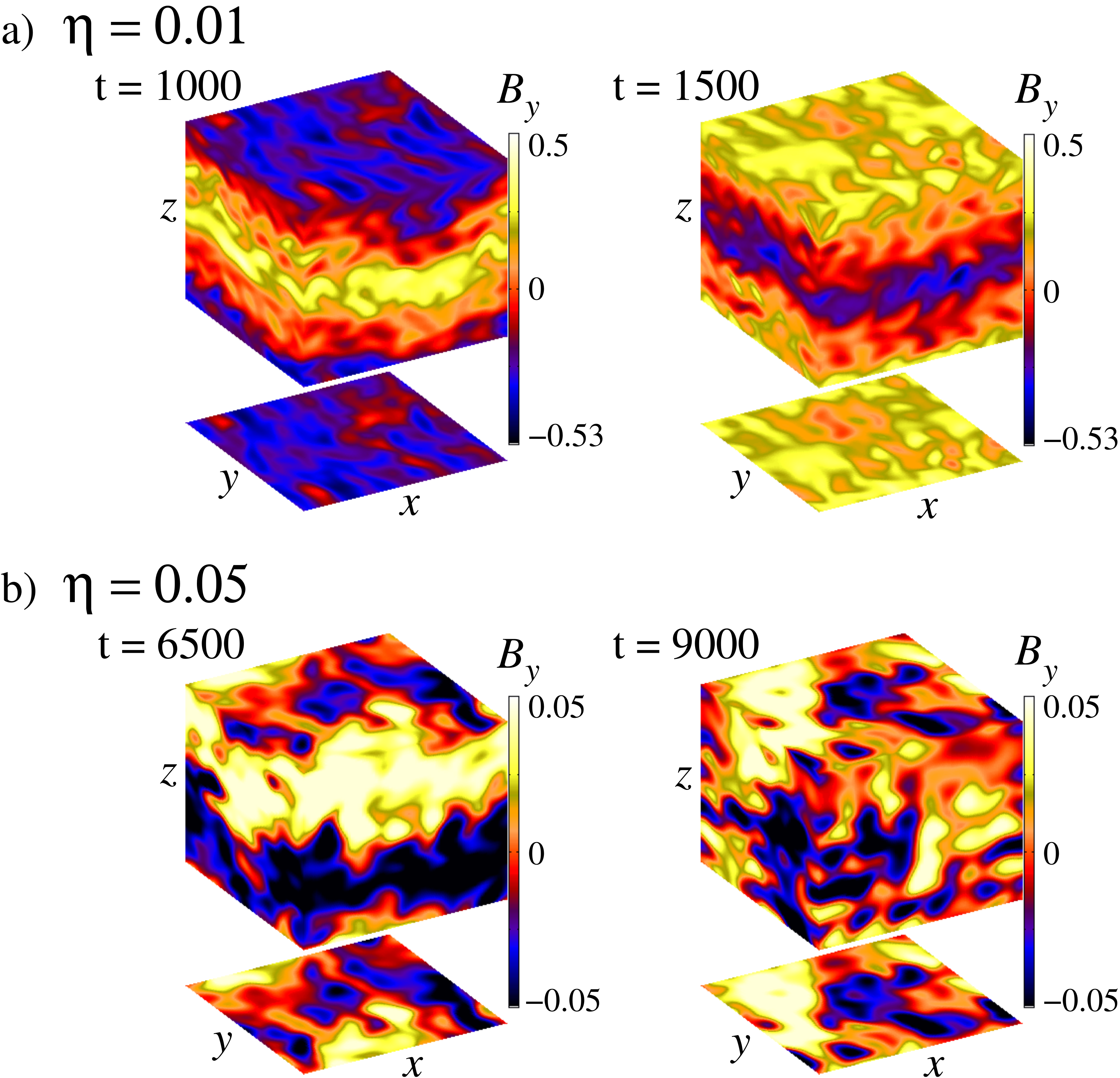}
 \caption{\label{fig patts} Intensity plots of $B_y$. {\bf (a)} 
The propagation of a large-scale coherent pattern along the $z$ direction for $\eta=0.01$. 
{\bf (b)} Switching between ordered $(t=6500)$ and disordered $(t=9000)$ patterns for $\eta=0.05$.}
 \end{figure}

We fix $\nu=0.005$, which in the absence of magnetic fields
corresponds to a weakly turbulent flow with $Re \approx 16$.
For large values of $\eta$, the seed magnetic field 
decays rapidly and there is no dynamo. 
After the  onset of dynamo action at  $\eta \approx 0.053$ ($Rm \approx 1.5$), 
the magnetic energy starts to grow at the 
expense of kinetic energy, until it saturates. 
Examples of magnetic structures are depicted in Fig. \ref{fig patts} for two values of $\eta$ and different times. For $\eta=0.01$
[Fig. \ref{fig patts}(a)] there is a coherent large--scale $B_y$ component that propagates along the $z$ direction. 
For $\eta=0.05$ [Fig. \ref{fig patts}(b)], the magnetic field displays an intermittent switching
between ordered $(t=6500)$ and disordered $(t=9000)$ large--scale structures. The scale--bars reveal that $B_y$ at $\eta=0.01$ is
one order of magnitude stronger than at $\eta=0.05$.

\begin{figure*}
 \includegraphics[width=1.5\columnwidth]{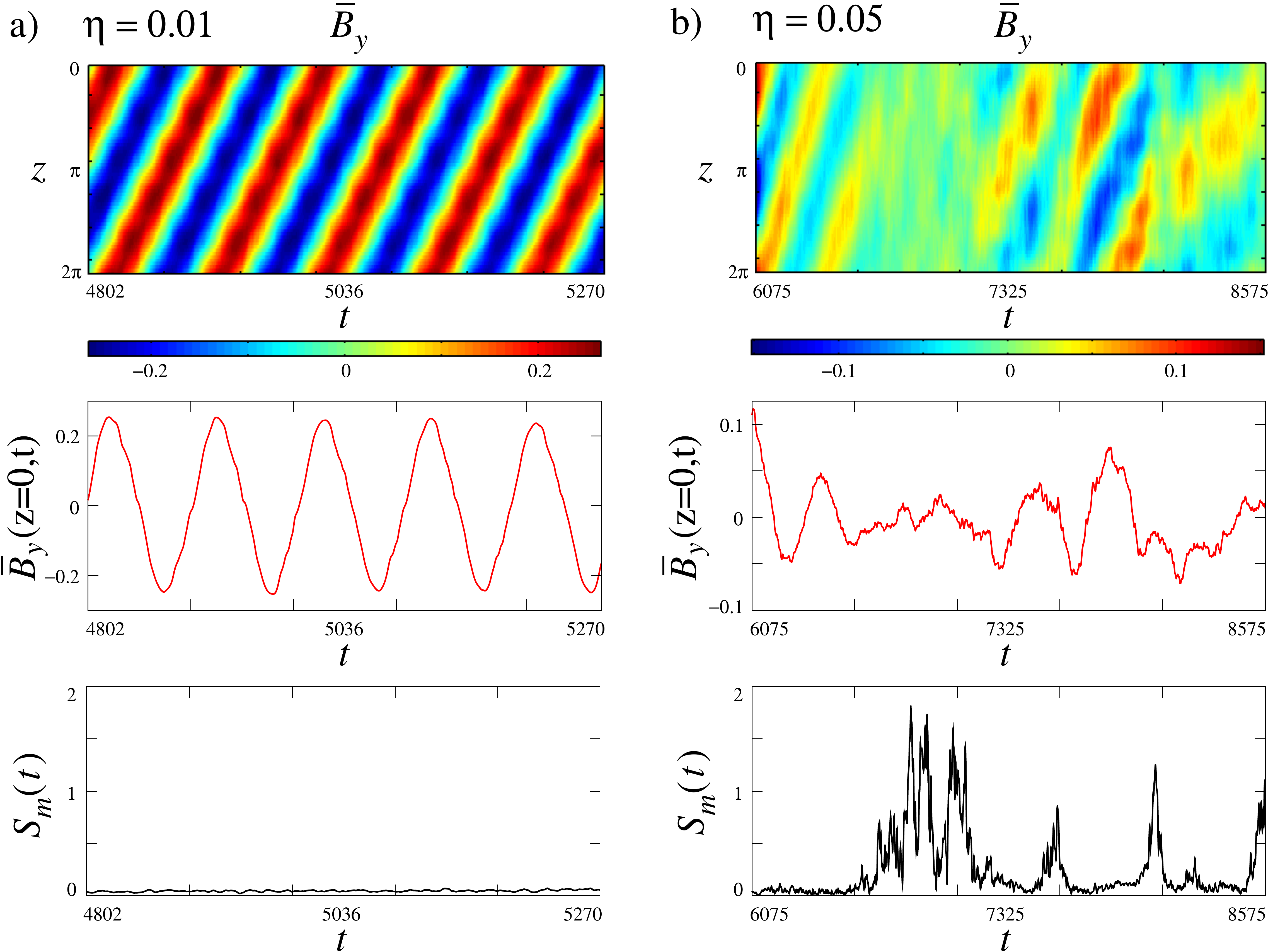}
 \caption{\label{fig xt} {\bf (a)} 
Space--time evolution of $\bar{B}_y$ (upper panel), the time series of $\bar{B}_y$ at $z=0$ (mid panel) 
and the spectral entropy $S_m(t)$ (lower panel) for $\eta=0.01$. 
{\bf (b)} Same as (a) but for $\eta=0.05$, displaying an intermittent dynamo.}
 \end{figure*}

A better understanding of the spatiotemporal dynamics can be obtained
by computing $\bar{B}_y$, the $xy$--averages of \textbf{\emph{B}}. 
The upper panel in Figure \ref{fig xt} shows the space--time evolution of $\bar{B}_y$
 and the mid panel shows 
the time series of $\bar{B}_y$ at the point $z=0$. The lower panel shows the spectral entropy $S_m(t)$, which is 
a measure of spatial complexity computed from
the power spectra of $\bar{B}_y$ following \citet{rempel09b} and \citet{chian10}. 
The left column refers to $\eta=0.01$ and the right column to $\eta=0.05$.
The left column shows that the mean--field for $\eta=0.01$ propagates like a robust spatially coherent dynamo wave.
The corresponding velocity field displays a mean flow with propagating oscillations. 
The direction of propagation is arbitrary and depends on the initial condition, which shows that there is multistability in the system.
For $\eta=0.05$ (right panel) the mean--field is more fragile and there is on--off intermittency, with phases of spatially disordered patterns interspersed with 
phases of spatially coherent structures. We call the 
regime at $\eta=0.01$ ``dynamo wave" and at $\eta=0.05$ ``intermittent dynamo".
The time--averaged values of the spectral entropy are $\left<S_m(t)\right>_t \approx 0.045$ for $\eta=0.01$ and $\left<S_m(t)\right>_t \approx 0.33$ for $\eta=0.05$.

The effect of the magnetic field on the velocity field and its transport properties can be studied using the maximum finite--time
Lyapunov exponent (FTLE). 
%
%
%
The maximum FTLE, $\sigma_1^{t_0+\tau}(\boldsymbol{x})$, gives the finite-time average of the maximum rate
of divergence or stretching between the trajectories of a fiducial
particle at $\boldsymbol{x}(t)$ and its neighboring particles from time $t=t_0$ to $t=t_0+\tau$ \citep{shadden05}. 
A positive $\sigma_{1}$ is the signature of chaotic streamlines in
the velocity field.
Finite-time Lyapunov exponents are able to detect 
Lagrangian coherent structures, which 
are the time--dependent analogous of stable and unstable manifolds
of invariant sets in time-independent velocity fields. 
For a three--dimensional time-dependent velocity field, regions of maximum material
stretching generate local maximizing curves ({\it ridges}) in the FTLE field. 
Thus, repelling LCS (finite-time stable manifolds) produce 
ridges in the maximum FTLE field in the forward--time dynamics  
and attracting LCS 
(finite-time unstable manifolds) produce ridges in backward--time
\citep{haller01,shadden05,padberg07}. 

Since backward--time integration of dissipative systems is a major problem due to numerical instabilities \citep{celani04},
we have to resort to interpolation of recorded data. A run 
from $t_0-\tau$ to $t_0+\tau$
is conducted and full three-dimensional snapshots of the velocity field are saved at each $dt=0.5$ time interval.
Linear interpolation in time and third--order Hermite interpolation in space are used to obtain
the continuous vector fields necessary to obtain the particle trajectories. For backward--time, the interpolated snapshots 
from $t_0$ to $t_0-\tau$ are used
 and the particle trajectories are computed with a fourth--order 
Runge--Kutta method. The choice of the spatial interpolation scheme may affect the local dynamics of individual particles, 
which can result in minor changes in the delimitation of some of the material lines detected. 
Here, we adopted a third order interpolation, which is the standard scheme employed in the 
literature for computing FTLE (see \citet{haller00,shadden05,padberg07,mathur07,mendoza10}). 
We compared the results obtained for the ABC flow with the known results from the literature and they show excellent agreement.

\begin{figure*}
 \includegraphics[width=1.2\columnwidth]{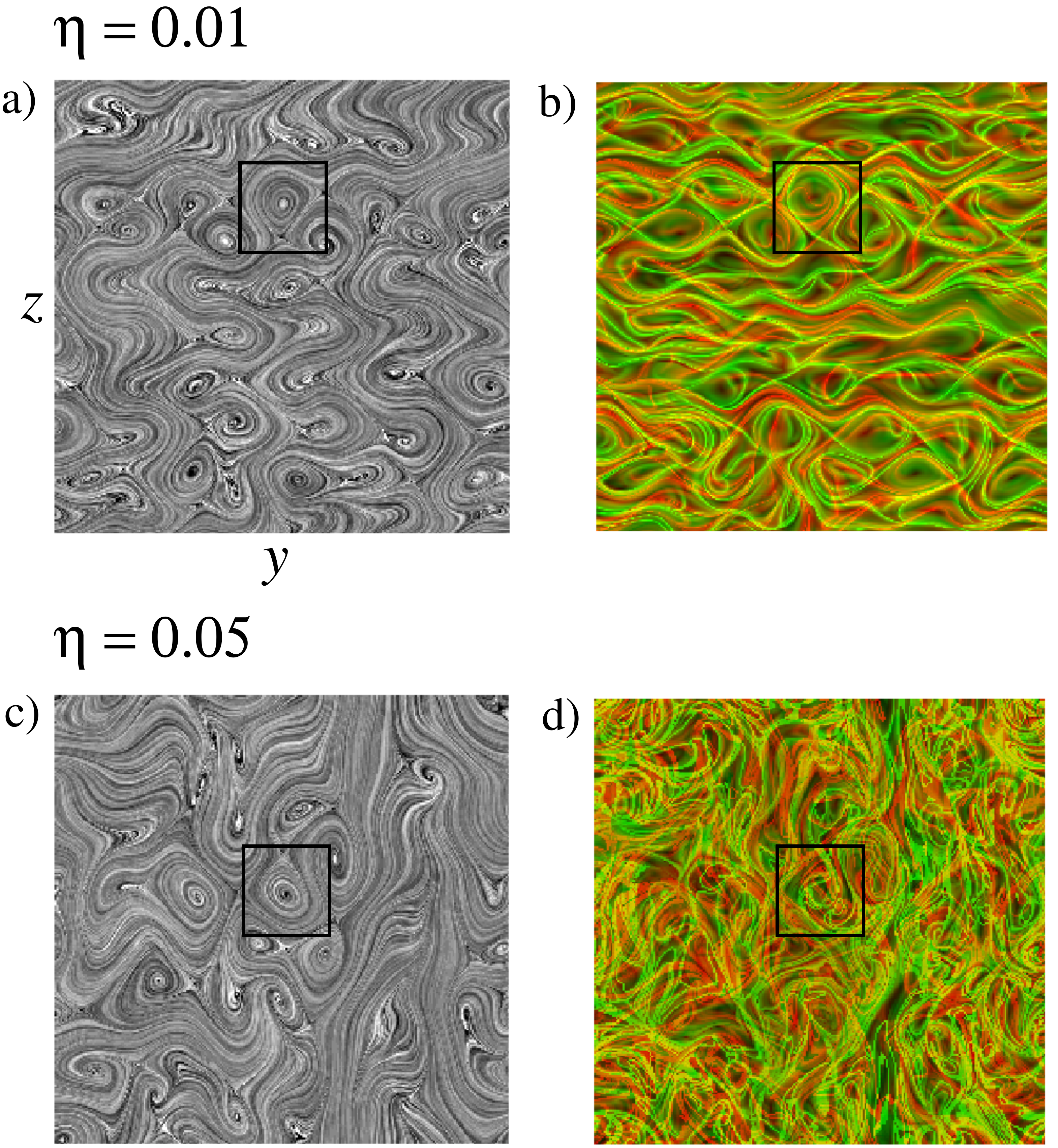}
 \caption{\label{fig manis} {\bf (a)} Line integral convolution plot of the $(y, z)$ components of 
a snapshot of the velocity field at $x=0$ for the wave 
dynamo at $\eta=0.01$. {\bf (b)} The corresponding repelling (green) and attracting (red)
LCS represented by material lines; {\bf (c)} and {\bf (d)}: same as (a) and (b), 
but for the intermittent dynamo at $\eta=0.05$.}
 \end{figure*}

\begin{figure*}
 \includegraphics[width=1.2\columnwidth]{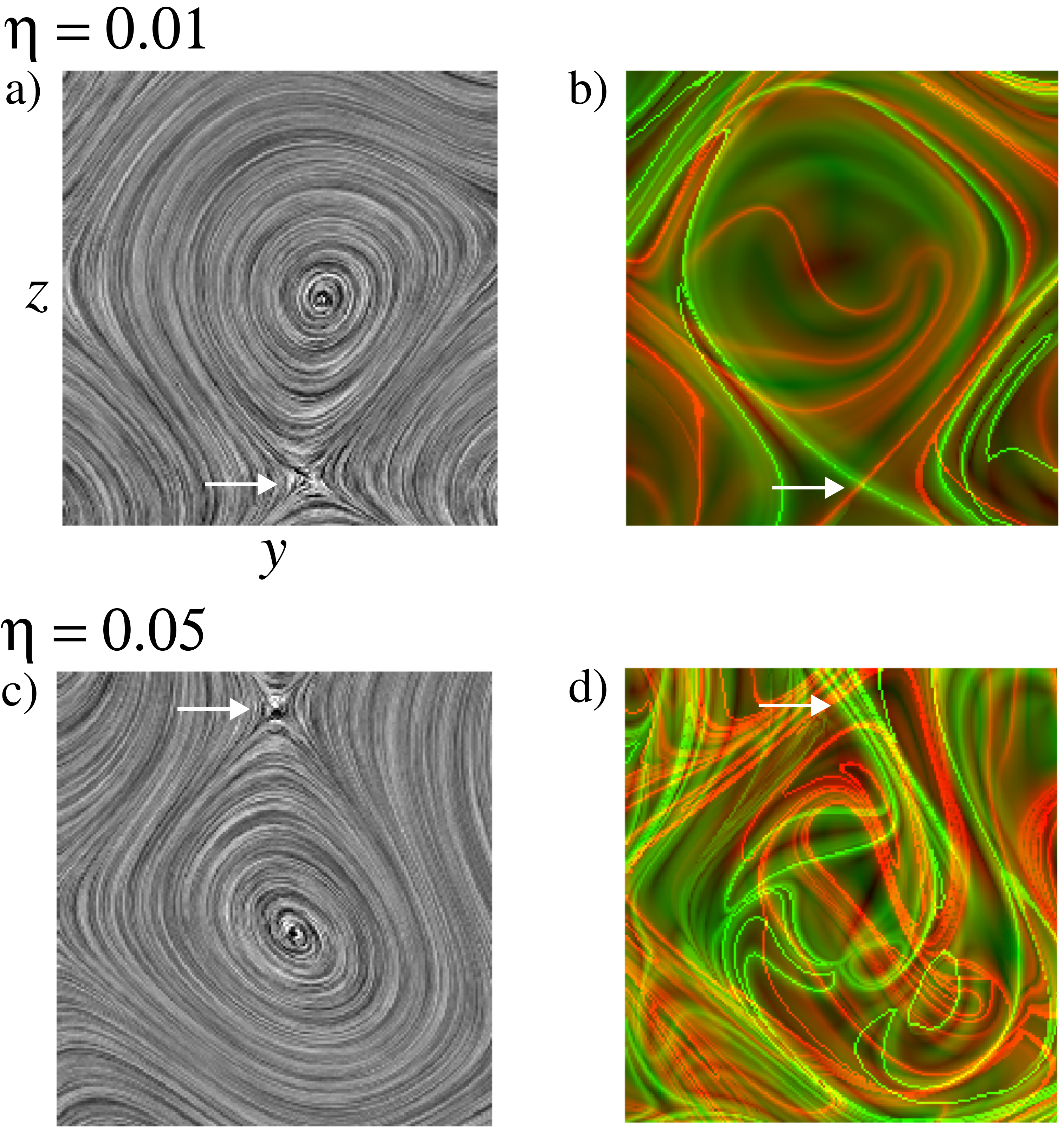}
 \caption{\label{fig zoom} Enlargement of the rectangular areas in Fig. \ref{fig manis}. {\bf (a)} and {\bf (b)}: the velocity field 
and Lagrangian coherent structures for $\eta=0.01$; {\bf (c)} and {\bf (d)}: same as (a) and (b) but for $\eta=0.05$.}
 \end{figure*}

Figure \ref{fig manis} illustrates the difference between the LCS of the wave and intermittent dynamos.
Figure \ref{fig manis}(a) is a visualization of the $(y,z)$ components of the velocity field for $\eta=0.01$ at $x=0$  
using the technique of line integral convolution,
which shows the integral curves of $(u_y, u_z)$ in different tones of gray. 
This snapshot was computed for $t_0=2000$, when the magnetic energy of the dynamo wave has already saturated (see Figure 2(a)). 
The corresponding LCS are shown in Fig. \ref{fig manis}(b), where green and red lines represent the repelling and
attracting material lines, respectively. The LCS represent the $\sigma_1^{t_0 + \tau}$  field computed with $\tau=\pm 10$.
Figure \ref{fig manis}(b) is plotted as a three-vector RGB image using Octave's imshow
routine, where the forward--time $\sigma_1$ field is stored in the ``green vector" and the backward--time $\sigma_1$ field
in the ``red vector". Notice that the intersections between high--intensity red and green lines may produce yellow points. 
We stress that the important feature
of these plots is not the absolute value of $\sigma_1$, but the ridges in its field, so the colormaps are normalized by
the largest value of $\sigma_1$.
Figures \ref{fig manis}(c)--(d) plot the velocity field and LCS, respectively, for the intermittent regime 
at $\eta=0.05$ ($t_0=2000$ and $\tau=\pm 10$). 
The LCS 
distinguish the dynamo wave
and intermittent dynamo regimes quite well. This becomes clearer in Fig. \ref{fig zoom}, which depicts enlargements
of the rectangular regions in Fig. \ref{fig manis}. 
For the dynamo wave [Figs. \ref{fig zoom}(a)--(b)], a large eddy in Fig. \ref{fig zoom}(a) is seen in the LCS plot 
of Fig. \ref{fig zoom}(b) as
a ``smooth" region with low level of particle dispersion bordered by attracting and repelling
material lines. 
The entanglement of attracting and repelling LCS is responsible for the transport of particles 
between eddies (in two-dimensional flows this transport mechanism is called lobe dynamics \citep{rom-kedar90}). 
The X--point marked with an arrow in Fig. \ref{fig zoom}(b) specifies 
the location of a hyperbolic trajectory nearby a point where the velocity field is instantaneously zero in the $(y,z)$ projection
[see the velocity field near the arrow in Fig. \ref{fig zoom}(a)]. 
The material lines are cross sections of material surfaces, and trajectories approach
the X--point along the green line and are repelled from it along the red line.
For the intermittent dynamo [Figs. \ref{fig zoom}(c)--(d)], the entanglement is much more complex, even 
though the eddies in Figs. \ref{fig zoom}(a) and \ref{fig zoom}(c) look similar. 
The Lagrangian plot unveils an intricacy of local structures which is not seen in the Eulerian frame.
The arrows in Figs. \ref{fig zoom}(c)--(d) mark
the location of an X--point whose time--dependent manifolds, or LCS, fill the phase space
in such a way that a border of the eddy cannot be identified. In the intermittent regime, the numerous crossings between
attracting and repelling material lines enhance transport between regions. This transport can
be quantified by the maximum finite--time Lyapunov exponent $\sigma_1^{t_0+\tau}$ 
of the particle trajectories. The mean value for the dynamo wave at $\eta=0.01$, obtained from a PDF computed with the trajectories
of $64^3$ particles evenly distributed 
in the box, is $\sigma_1^{t_0+\tau} \approx 0.25$. 
For the intermittent dynamo at 
$\eta=0.05$, $\sigma_1^{t_0+\tau} \approx 0.32$
and, therefore, the resulting chaotic mixing is more efficient. Although the LCSs vary according
to $t_0$, in general the LCS fields computed at $\eta=0.05$ display higher degree of complexity than at
$\eta=0.01$.


The enhancement in the flow's chaoticity when the magnetic diffusivity is increased from $\eta=0.01$ to $\eta=0.05$
is the result of a reduction of the effect of the Lorentz force upon the velocity
field. For $\eta=0.01$, $B_{\rm rms} \approx 0.27$ and $u_{\rm rms} \approx 0.29$, so the magnetic field can become strong enough to suppress 
Lagrangian chaos in the velocity field, inhibiting particle transport. As a result of lower chaoticity,
as well as the backward transfer of magnetic energy from small to large scales due to kinetic helicity ($\alpha$--effect) 
present in this system, 
the magnetic field \textbf{\emph{B}} saturates in an ordered state with the scale of the box and the 
mean--field dynamics resembles a spatially--coherent propagating wave [Fig. \ref{fig xt}(a)]. 
For $\eta=0.05$, stronger magnetic diffusivity causes the magnetic field to be damped, with $B_{\rm rms} \approx 0.076$ and $u_{\rm rms} \approx 0.38$.
A weaker magnetic field has small impact on the velocity field and Lagrangian chaos 
becomes stronger, with enhanced particle transport and chaotic mixing. The chaotic motions
of the flow carry the magnetic field lines and generate the disordered \textbf{\emph{B}} field states. On the other hand,
stronger chaos in the velocity field leads to enhanced stretching, twisting and folding of magnetic field lines, which tends to cause the magnetic
energy to grow \citep{childress95}. The growth of \textbf{\emph{B}} backreacts on the velocity field, suppressing chaoticity
again and leading to the intermittent occurrence of ordered \textbf{\emph{B}} field patterns observed in 
the intermittent dynamo of Figs. \ref{fig patts}(b) and \ref{fig xt}(b). 

\section{Nonlinear dynamo saturation in the B2001 flow}

\begin{figure*}
 \includegraphics[width=1.4\columnwidth]{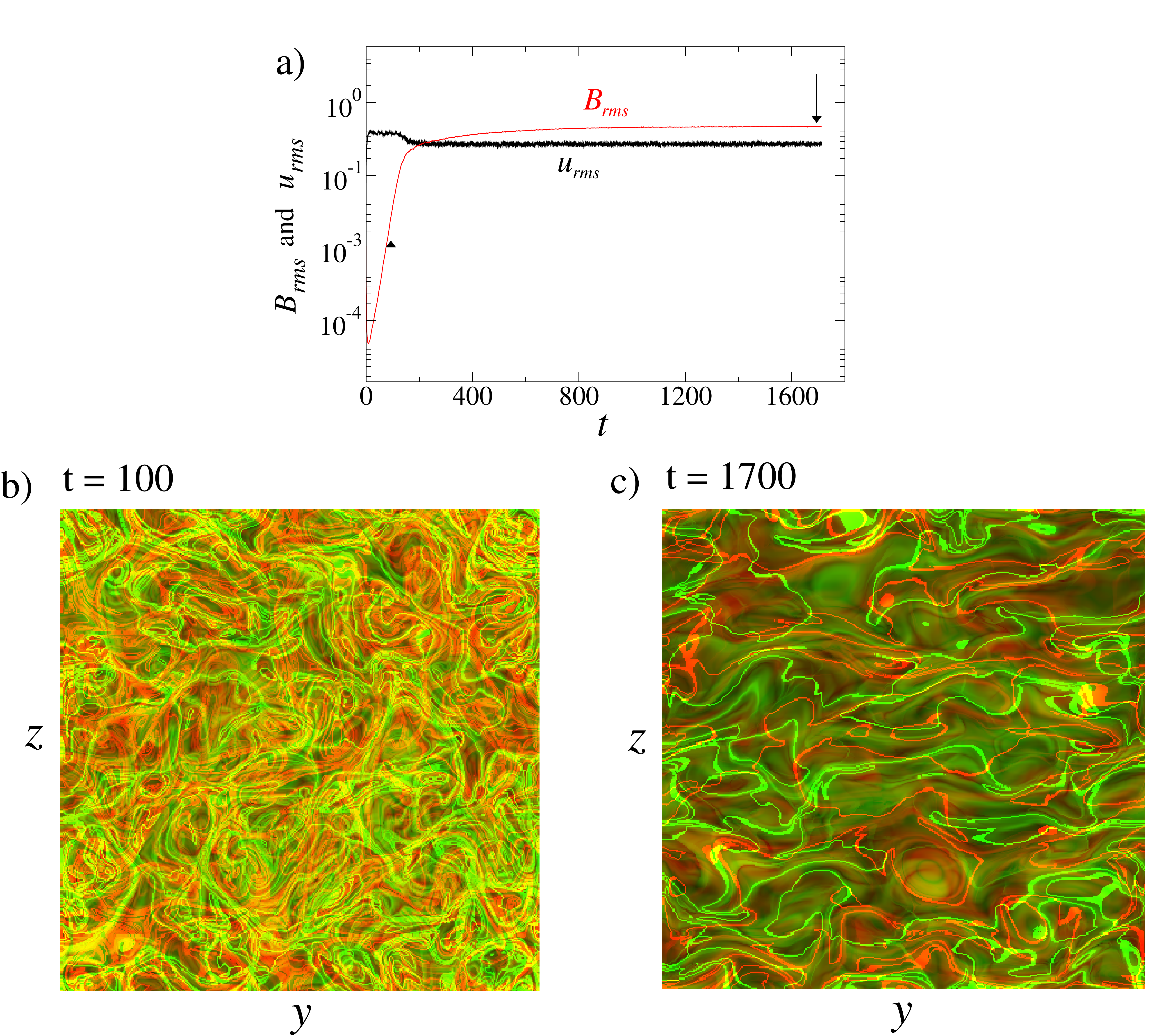}
 \caption{\label{fig b2001} Turbulent simulations with random helical forcing. {\bf (a)} Evolution of the total rms magnetic (red) and velocity (black)
 fields. The arrows point to $t=100$ and $t=1700$; {\bf (b)} LCS for the kinematic dynamo at $t=100$; (c) LCS for the saturated dynamo at $t=1700$.}
 \end{figure*}

For this section we have performed computations
described in \citet{axel2001} (hereafter, B2001), where the MHD equations are solved with a helical forcing 
with a time--dependent wavevector. At each time step, there is a random choice of wavevector with 
wavenumber $k_f$ around 5. The resulting flow is essentially the prototype of the $\alpha^2$ dynamo
of mean--field dynamo theory. We have adopted
run 3 of B2001, where $\nu=\eta=0.002$ and the Reynolds number based on the box size is about 600 
($Re = Rm = 18$ as defined in Section 2 above)
for a numerical resolution of $128^3$. 
Due to an inverse cascade of magnetic helicity discussed in B2001, the magnetic field develops 
a robust spatially coherent mean--field pattern similar to the case $\eta=0.01$ above [Fig. \ref{fig patts}(a)].
Figure \ref{fig b2001}(a) shows the time--evolution of $u_{\rm rms}$ and $B_{\rm rms}$ in log--linear scale, revealing
the exponential growth of $B_{\rm rms}$ in the kinematic phase, before saturation. Figures \ref{fig b2001}(a),(b)
show the LCS computed for $\tau=\pm 10$ at $t_0=100$ (kinematic dynamo) and $t_0=1700$ (saturated nonlinear dynamo), respectively.
In the kinematic regime the patterns of material lines in the LCS plot
are highly complex, and the chaotic tangle permeates the phase space, which favors the growth of magnetic energy.
In the nonlinear regime, since $B_{\rm rms}$ becomes considerably higher than $u_{\rm rms}$ (superequipartition), 
the chaoticity of the velocity field is
strongly decreased due to the Lorentz force \citep{cattaneo96,zienicke98}. 
The crossings between the main attracting and repelling lines are scarce, so there is 
comparatively little dispersion of passive scalars and transport is inhibited. 
The level of chaotic mixing quantified by the average FTLE is $\sigma_1^{t_0+\tau} \approx 0.34$ for the kinematic
dynamo and $\sigma_1^{t_0+\tau} \approx 0.18$ for the saturated dynamo.

\section{Conclusions}

We have shown that Lagrangian coherent structures (LCS) can be used for an in--depth exploration of particle transport in
3--D MHD dynamo simulations. Our results agree with the previous results by \citet{cattaneo96} and \citet{zienicke98},
who showed that the modification of the velocity field due to stronger \textbf{\emph{B}} becomes clearer 
by examining the Lagrangian properties of the flow as measured by the finite--time Lyapunov exponents (FTLE).  
Here, in addition to computing the forward--time FTLE field, the detection of attracting material lines as
ridges in the backward--time FTLE field provides the pathways that are more likely to be followed by passive scalars in the fluid.
Moreover, the superposed plots of both attracting and repelling LCS permit the identification of the 
principal mixing zones of the fluid.

The two dynamo models adopted in this work exhibit weak turbulence, with reasonably low Reynolds numbers. Our
goal was not to present state--of--the--art numerical simulations, but to introduce the LCS technique
in the context of space/astrophysical plasmas using two important topics in the theory of nonlinear dynamos,
namely, the onset of intermittency and the nonlinear saturation of the magnetic energy.
Regarding the first topic, the connection between the Lorentz force and on--off dynamo intermittency in ABC flows was   
studied by \citet{alexakis08}; on--off intermittency has also been observed in laboratory experiments with a dynamo 
generated by a flow of liquid sodium \citep{ravelet08,monchaux09}; besides, intermittent chaotic dynamos have been suggested as 
the cause of the long periods of low solar activity in the solar cycle, known as grand minima \citep{spiegel09}. 
Concerning the second topic, it is one of the fundamental questions in dynamo theory and 
there is an extended list of papers that discuss the nonlinear saturation of \textbf{\emph{B}} 
in dynamo simulations in periodic boxes with moderate Reynolds numbers \citep{axel95,cattaneo96,zienicke98,axel2001,axelPR,kapyla09,cattaneo09}.

The LCS method can be readily employed in a number of problems related to the turbulent transport of passive scalars, including 
observational data, provided an estimation of the velocity vector field is available. Such estimations can be obtained 
from digital images using techniques such as the optical flow algorithm, employed by \citet{colaninno06} to
extract the velocity field from images of coronal mass ejections obtained with the SOHO LASCO C2 coronagraph. 
Solar subsurface flows can also be inferred from helioseismic data \citep{woodard02}, thus LCS
can aid the tracing of particle transport by turbulence in stellar interiors. 

In conclusion, a proper understanding of Lagrangian chaotic mixing is crucial for
understanding the dynamics of nonlinear dynamos
as well as the elaboration of models of stellar interiors that can correctly account for the turbulent transport of particles, energy and momentum
in convective zones, and we believe LCS are an innovative tool that should be further explored in Astrophysics.



\acknowledgments

ELR thanks the University of Cambridge for the hospitality and 
Prof. M. Proctor for valuable discussions.
ACLC acknowledges the award of a Guggenheim Fellowship and the hospitality of Caltech.
ELR and ACLC acknowledge the support by CNPq (Brazil) and FAPESP (Brazil).

\clearpage

\end{document}